# Search, Structure, and Sentiment: A Comparative Analysis of Network Opinion in Different Query Types on Twitter

**Joshua Midha**

West Windsor-Plainsboro High School South

West Windsor, NJ, USA

jmidha05@gmail.com

December 25, 2022

## Abstract

Understanding the relationship between structure and sentiment is essential in highlighting future operations with online social networks. More specifically, within popular conversation on Twitter. This paper provides a development on the relationship between the two variables: structure, defined as the composition of a directed network, and sentiment, a quantified value of the positive/negative connotations of a conversation. We highlight thread sentiment to be inversely proportional to the strength and connectivity of a network. The second portion of this paper highlights differences in query types, specifically how the aforementioned behavior differs within four key query types. This paper focuses on topical, event-based, geographic, and individual queries as orientations which have differing behavior. Using cross-query analysis, we see that the relationship between structure and sentiment, though still inversely proportional, differs greatly across query types. We find this relationship to be the most clear within the individual queries and the least prevalent within the event-based queries. This paper provides a sociological progression in our understanding of opinion and networks, while providing a methodological advancement for future studies on similar subjects.

## 1. Introduction

Conversation and threads of responses on online forums have recently emerged as an effective subject for researchers interested in diffusion, polarization, and discourse. Studies have explored the human behavioral aspects (Fischer et al., 2011; Buccafurri et al., 2015; Gao et al., 2012), socio-physical consequences (Serrano et al., 2016; Liu et al., 2011; Ariyaratne 2016), and political systems changes (Effing et al., 2011; Vergeer, 2015; Tumasjan et al., 2011). Through these analyses, the use of social network analysis (SNA) and sentiment analysis has become a commonality. Varieties of studies have charted interaction and used natural language processing to study polarization. Amidst this, institutions and researchers alike have found patterns within these forums. As such developing theories of cascading tweets (Tsuwaga, 2019), echo chambers (Smith et al., 2014), and network structure (Feller et al., 2011). These theories and in-depth analyses provided by investigative journalists have provided a better understanding on the role of online communications within social systems.

Two growingly important and often compared features of the online communication studied are structure and sentiment. Structure discusses the physical attributes of a network and the links between its agents. The rudimentary building blocks of an online social network are its nodes and links. A network's structure consists of its size which is modeled by the number of agents, complexity that is given by degrees of connectivity and the links, and homogeneity which is shaped by the variety of the aforementioned factors across the whole network (Burt, 1980). With these three notions, a structure of a conversation—small or big, online or in-person—can be modeled. Sentiment, in this context, is a metric of emotion provided to help evaluate text. In an online social network, sentiment provokes conversation and develops an identity of a network. Ranging from a polarized echo chamber to mundane conversations, sentiment is ever present and dynamic. A network, in turn, is made by the agents and their formation, and the emotional values which guide it.

The connection between structure and sentiment is of interest to several researchers. A potential correlation between the two could help with the detection of polarized communities or echo chambers early on. This practice could help prevent opinion terrorism and in turn increase the safety of online platforms. In order to do so, some papers have provided insights on how structure may influence sentiment (Boutyline et al., 2017). However, there has been a lack of focus in literature to differentiate factors which may provide conflicting theories of correlation.

Of those factors, this paper is specifically interested in variation of the query types. The type of query can determine the type of pattern returned. For example, the structure of a network in regards to a broad topic will vary significantly from the structure of a network discussing a small sized rural town (Packer et al., 2012). This paper argues that while most researchers conclude based on a topic, geocode, event or person as their query, a comparative perspective on all is necessary to root out potential differences.

This paper aims to provide a comparative analysis of the correlation between structure and sentiment in different subject types. It differentiates between a topical query (3.1), an event query (3.2), a geographic query (3.3), and a query on individuals (3.4). The paper's research question are as follows:
1. Is there a notable difference within the structure of each query type?
2. Is there a notable difference within the sentiment of each query type?
3. Is there a notable correlation between sentiment and structure for each of the query types?
4. Is there a notable difference within the patterns between sentiment and structure for each query type?
5. How can the differences within query types help inform future research on social network structure and community sentiment?

## 2. Methodology

This paper addresses potential differences through studying conversations on Twitter. Related works have used Twitter as a laboratory for understanding patterns on diffusion and social change (Yang et al., 2010; Taxidou et al., 2014) and the robust Twitter API allows for large quantities of data to be pulled and analyzed. The methodology proposed by this paper uses the Twitter API to pull tweets from the platform and model them using R. Within R, we make use of the rTweet (Kearney et al., 2022), iGraph (Nepusz, 2022), Syuzhet (Jockers, 2020), and other base packages to model and chart structure and calculate sentiment.

First, tweets are queried from the API. The terms of this query differ as per each condition tested, but 950 tweets are pulled per query. Topics for the queries are determined using the United States popular search items for each category. For example, if we are launching a query for an individual query, we will randomly select one of the individuals from the list of popular individual queries on Twitter. The search results are then specified using the "mixed" type within the Twitter Rest API. This allows us to gather a dataset which encompasses more conversations with direction than the "popular" or "recent" strings would bear. A dataframe is then constructed using the tweets pulled and the text is extracted from each tweet. After cleaning the text and using the "get_sentiment" function of the Syuzhet package, a sentiment value, where a more positive number signifies a positive text and the inverse for negative numbers, is returned. We then average the sentiment value of each tweet in order to get an average for the megathread, or larger conversation. This process is then rerun for 100 duplicate queries to get a more precise average value for the conversation. We will represent this value with α.

As for the structure, we use the "network_graph" function from the rTweet package to create an iGraph object which represents the network of the dataframe. Due to the source and target classifications provided within the rTweet query's dataframe, the network_graph function is able to effectively create a directed network. This object is then plotted and the sentiment value is attached for accessibility of comparison. Below is an example.

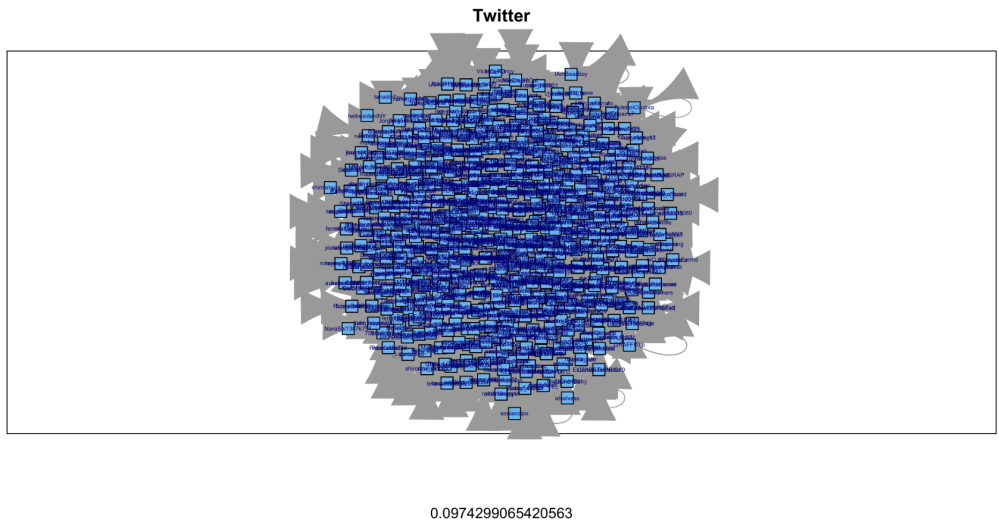

**Figure 1**: Example network highlighting structure and sentiment of topical query "Twitter"

To help highlight fragmentation of each subject and quantify the structure, we attach a compilation of the component distributions. Using the connectivity of the iGraph object through two depth searches, we can note the strong components and the weak components. The weak components resemble a less reliable, more temporary community while the strong highlights the inverse (Wu et al., 2007). We will do this for each iteration of each query and calculate an average strong and weak count which are going to be rounded to the nearest whole number. In our context, this resembles the clusters we are interested in and our goal is to see a correlation between the ratio of weak to strong and the sentiment. We will calculate a ratio, ß, which shows the presence of strong to weak communities within the discussion thread. Thus highlighting how close knit the community is.

The null hypothesis states that there is no correlation between ß and α. The alternative hypothesis pushed forth by the paper argues that the more close knit a community is, the less polarizing its sentiment will be. This can be explained by ideas in online and in-person communities where response to change is better expected, tolerated, and minimized by long standing communities versus temporary ones (Qu et al., 2009; Veysey et al., 2016). The second null hypothesis highlights no difference between each of the correlations. However, the second alternative hypothesis, $H_b$, argues for a difference in each of the r values for each query type. In the $H_b$, each query type is denoted by its leading letter. For example the correlation coefficient for the topical query is going to be $r_t$.

$$H_0 : r = 0$$
$$H_a : r = -1$$
$$h_0 : r_t = r_e = r_g = r_i$$
$$H_b : r_t \neq r_e \neq r_g \neq r_i$$

## 3. Results

Data for each query type was collected, visualized and included under the following sections.

### 3.1 Topical Queries

In a topical query, we launch a search for conversations on a certain topic. Using the Twitter API, we pull a series of trending topics and then run them using the aforementioned script. Below is a comparative chart of topical queries which follow the format of Figure 1.

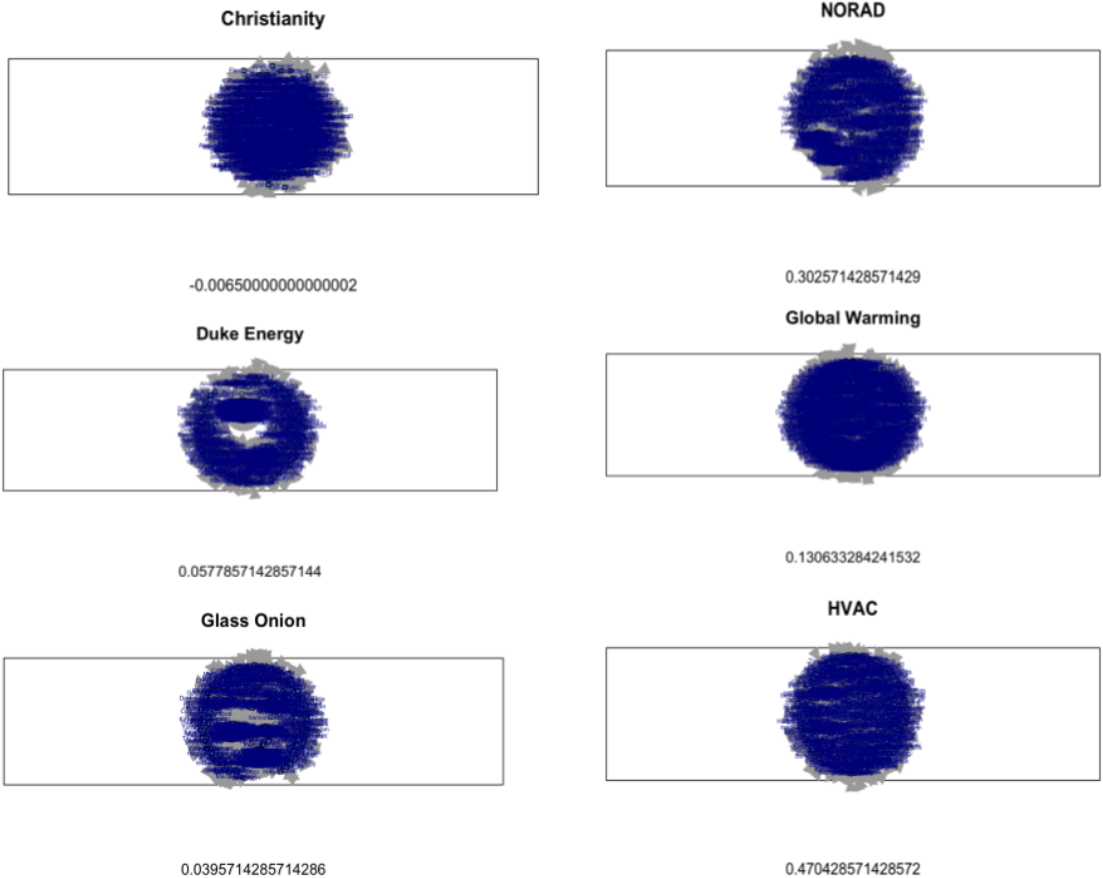

**Figure 2**: Six randomly selected subjects from the trending topics for the "Topical Query." A comparison of all six subjects and average sentiment highlighted from 100 interactions of each query.

As aforementioned in our methods, we are highlighting our component distribution and sustaining a further inquiry of structure through iGraph connectivity. Below is a table of the distribution and ratio of interest.

| Subject | Strong Count | Weak Count | Ratio (ß) | Sentiment (α) |
|---|---|---|---|---|
| Christianity | 336 | 245 | 0.7291666667 | -0.0065 |
| NORAD | 507 | 118 | 0.2327416174 | 0.3025 |
| Duke Energy | 164 | 97 | 0.5914634146 | 0.0577 |
| Global Warming | 603 | 156 | 0.2587064677 | 0.1306 |
| Glass Onion | 182 | 150 | 0.8241758242 | 0.0395 |
| HVAC | 516 | 142 | 0.2751937984 | 0.4704 |

**Table 1**: A count on distribution, both weak and strong, and ß, the ratio of strong to weak components, are included. The average sentiment from all query iterations is also attached for ease of comparison.

## 3.2 Event-Based Queries

For the event-based queries, we launch a search based on a specific event. Whether it be a day or a period in time, there are a series of tweets which use either Hashtag format or repetitive mentioning of these events within the text. Below is a comparative chart of event-based queries.

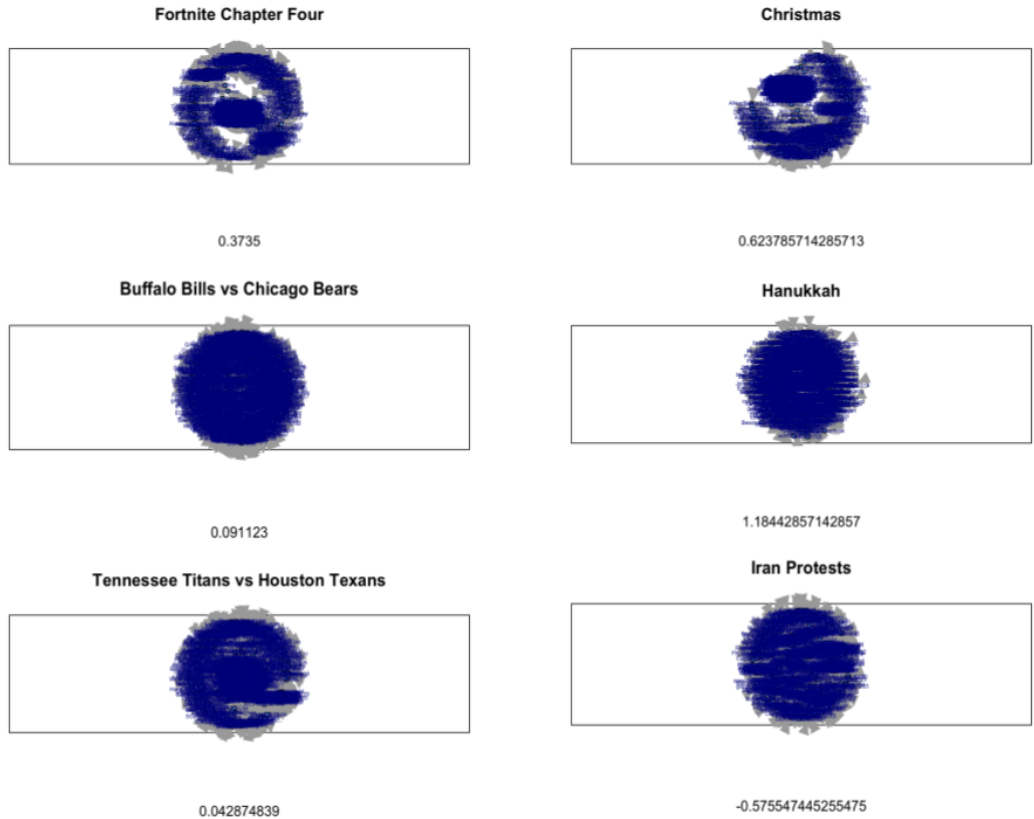

**Figure 3**: Six randomly selected subjects from the trending topics for the "Event-Based Query." A comparison of all six subjects and average sentiment highlighted from 100 interactions of each query.

For some events, specifically the American football matches, Boolean operators and a mixed format query was launched to gather data which discussed the event and not the specific teams. The component distribution for each of the six subjects is presented in the table below.

| Subject | Strong Count | Weak Count | Ratio (ß) | Sentiment (α) |
|---|---|---|---|---|
| Fortnite | 558 | 66 | 0.1182795699 | 0.3735 |
| Christmas | 633 | 102 | 0.1611374408 | 0.6237 |
| Bills vs Bears | 652 | 511 | 0.7837423313 | 0.0911 |
| Hanukkah | 669 | 98 | 0.1464872945 | 1.1844 |
| Titans vs Texans | 483 | 301 | 0.6231884058 | 0.0429 |
| Iran Protests | 211 | 42 | 0.1990521327 | -0.5756 |

**Table 2**: A count on distribution, both weak and strong, and ß, the ratio of strong to weak components, are included. The average sentiment from all query iterations is also attached for ease of comparison.

### 3.3 Geographic Queries

Geographic queries are search methods which make use of coordinates, geocodes, location titles or any formatable parameter which returns the result for a specific location. Within the searches for each of these locations, and only if needed, we specified the radius to be based on past geocoding queries and sample data for each city. Additionally, all popular geographies selected were limited to urban areas. The comparative chart is below.

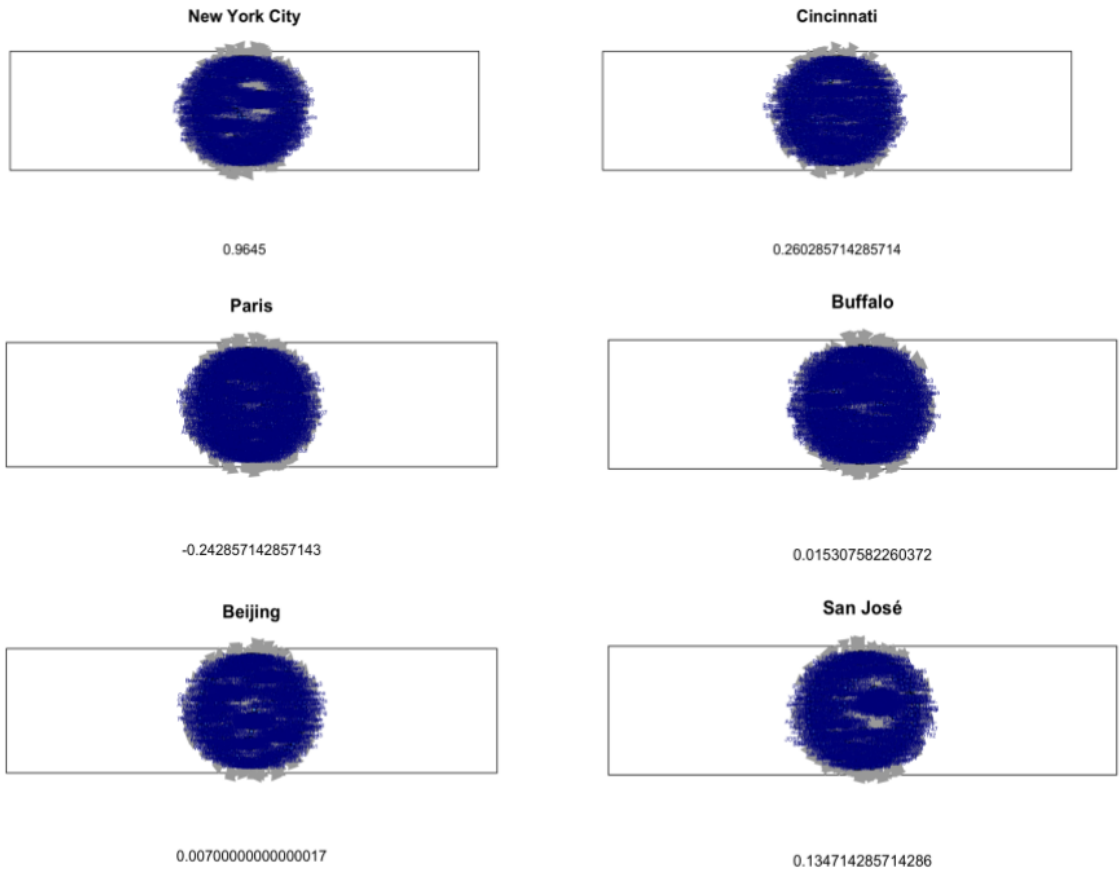

**Figure 4**: Six randomly selected subjects from the trending topics for the "Geographic Query." A comparison of all six subjects and average sentiment highlighted from 100 interactions of each query.

The component distribution is included below. In the trending lists, there was overlap between cities, coordinates, geocodes and zip codes. These were merged using the Boolean search operators to represent the same location.

| Subject | Strong Count | Weak Count | Ratio (ß) | Sentiment (α) |
|---------|--------------|------------|-----------|----------------|
| NYC | 340 | 27 | 0.07941176471 | 0.9645 |

| Cincinnati | 372 | 108 | 0.2903225806 | 0.2602 |
| Paris | 362 | 113 | 0.3121546961 | -0.2429 |
| Buffalo | 430 | 382 | 0.888372093 | 0.0153 |
| Beijing | 239 | 179 | 0.7489539749 | 0.0071 |
| San José | 295 | 68 | 0.2305084746 | 0.1347 |

**Table 3**: A count on distribution, both weak and strong, and ß, the ratio of strong to weak components, are included. The average sentiment from all query iterations is also attached for ease of comparison.

### 3.4 Individual-Based Queries

Individual queries use search terms from the trending lists which involve the discussion of one person. This person is the subject focus for the conversation and only tweets relevant to the subject are included in the conversation. Within these search terms, often merely searching the name of the individual of interest does not yield a varied dataset. As such, the query is shaped with Boolean search operators which include nicknames, alternative names, and other common sayings in order to capture all conversation regarding the subject. Below is a comparative chart on each query.

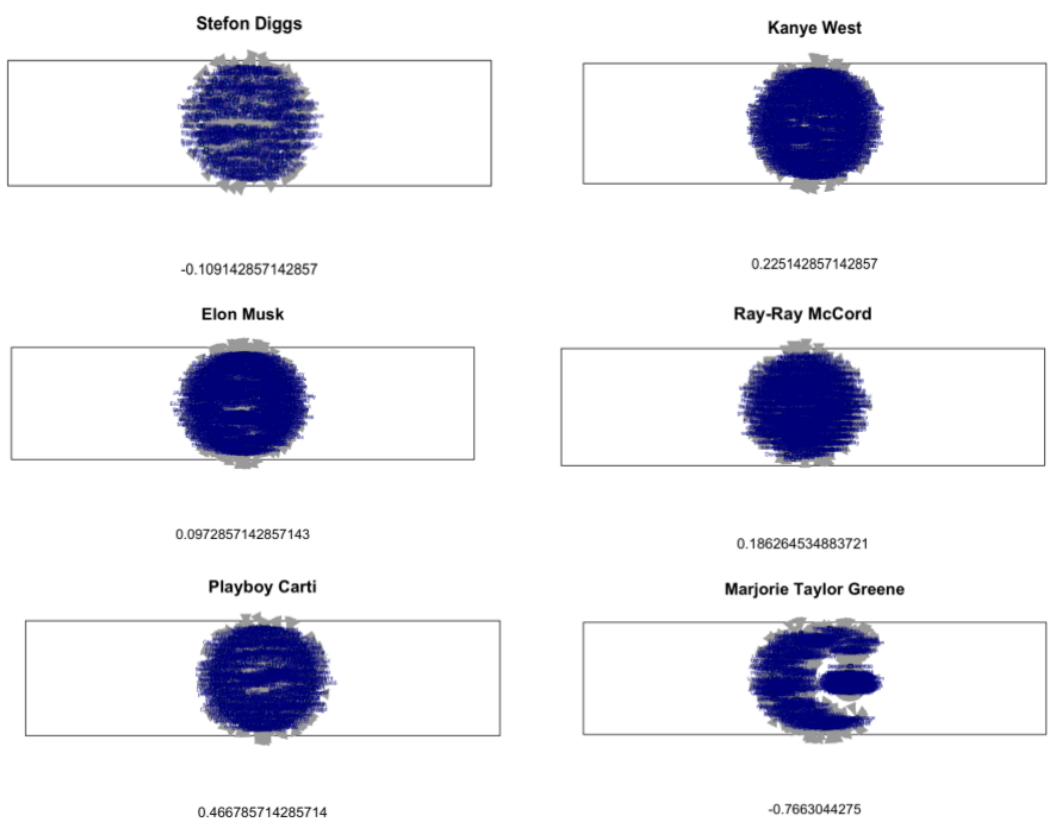

**Figure 5**: Six randomly selected subjects from the trending topics for the "Individual Query." A comparison of all six subjects and average sentiment highlighted from 100 interactions of each query.

Component distributions are included below. The table highlights the merged subjects and the distribution for each of them.

| Subject | Strong Count | Weak Count | Ratio (ß) | Sentiment (α) |
|---|---|---|---|---|
| Stefon Diggs | 129 | 65 | 0.503875969 | -0.1091 |
| Kanye West | 542 | 174 | 0.3210332103 | 0.2251 |
| Elon Musk | 361 | 215 | 0.595567867 | 0.0972 |
| Ray Ray McCloud | 702 | 271 | 0.386039886 | 0.1863 |
| Playboi Carti | 662 | 112 | 0.16918429 | 0.4668 |
| Marjorie Taylor Greene | 137 | 121 | 0.8832116788 | -0.7663 |

**Table 4**: A count on distribution, both weak and strong, and ß, the ratio of strong to weak components, are included. The average sentiment from all query iterations is also attached for ease of comparison.

## 4 Discussion

Provided are analyses of the results for each data type and an analysis of the cross-query data. These analyses are in regards to the aforementioned hypothesis.

### 4.1 Pearson Correlation and Analysis of Within-Query Data

Using the Pearson-Correlation formula, we can analyze the correlation between ß and α for each query type. The formula follows below and the "ggpubr" package was used for calculation and visualization (Kassambara, 2022).

$$r = \frac{\sum (x - m_x)(y - m_y)}{\sqrt{\sum (x - m_x)^2 \sum (y - m_y)^2}}$$

The correlation between ß and α for the Topical Query leaned towards the alternative hypothesis, $H_a$. However, since $p < 0.01$, it is not significant enough to prove the alternative.

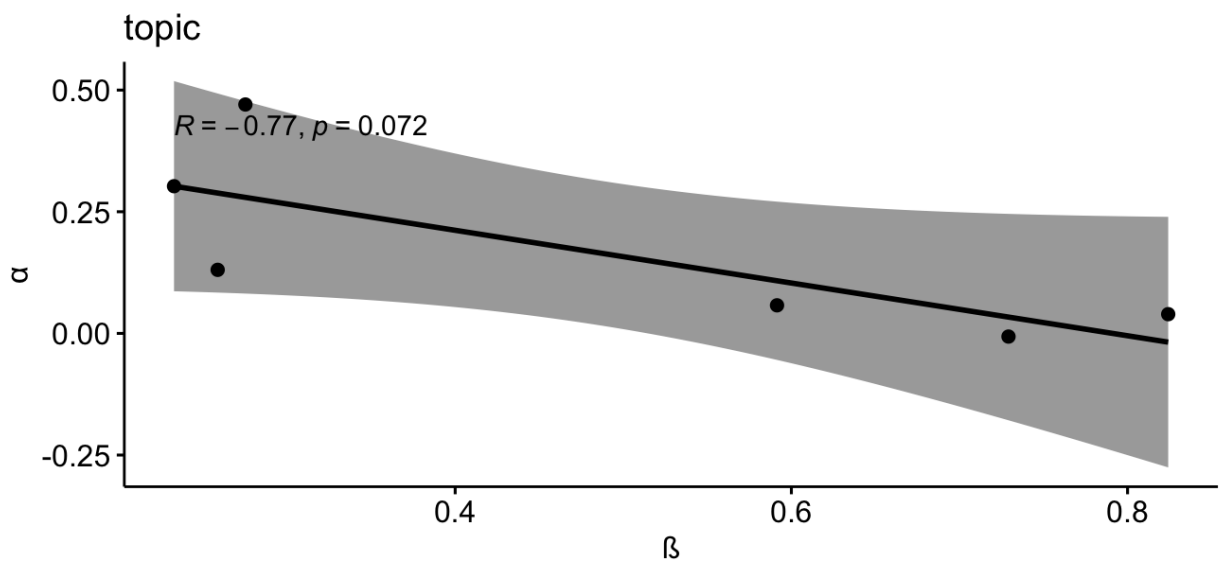

**Figure 6**: Using the ggscatter function, a correlation between the strong-weak ratio and the sentiment is highlighted. It has a correlation coefficient of -0.77 and a p-value of 0.072.

Similarly, the correlation between ß and α for the Event-Based Query leaned towards the alternative hypothesis, $H_a$. However, it did so to a significantly lesser degree. Again, since $p < 0.01$, it is not significant enough to prove the alternative.

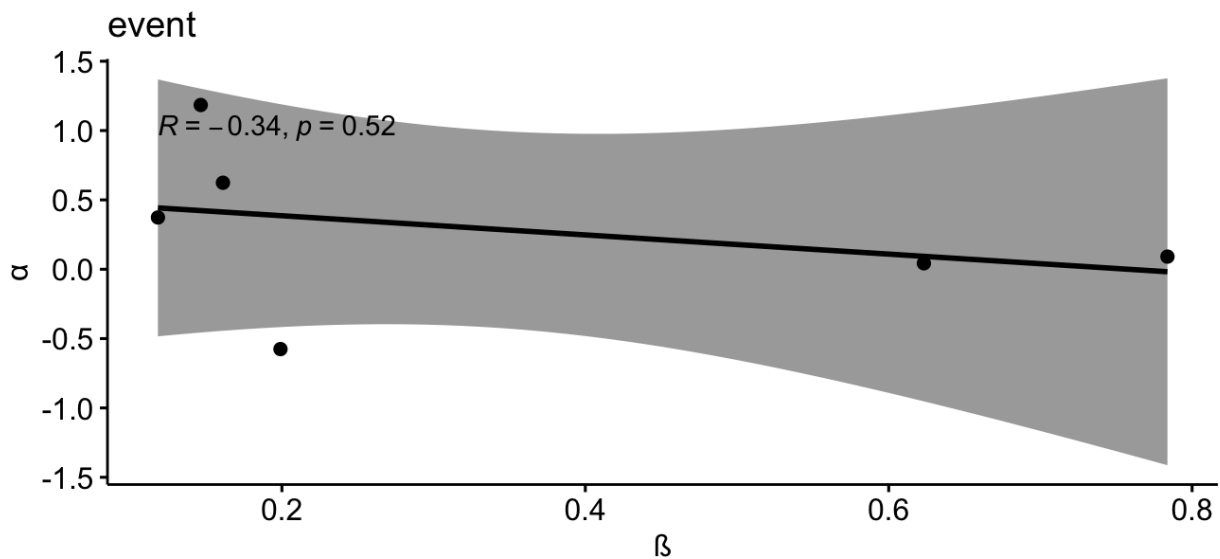

**Figure 7**: Using the ggscatter function, a correlation between the strong-weak ratio and the sentiment is highlighted. It has a correlation coefficient of -0.34 and a p-value of 0.52.

The correlation between ß and α for the Geographic Query leaned towards the alternative hypothesis, $H_a$. However, since $p < 0.01$, it is not significant enough to prove the alternative.

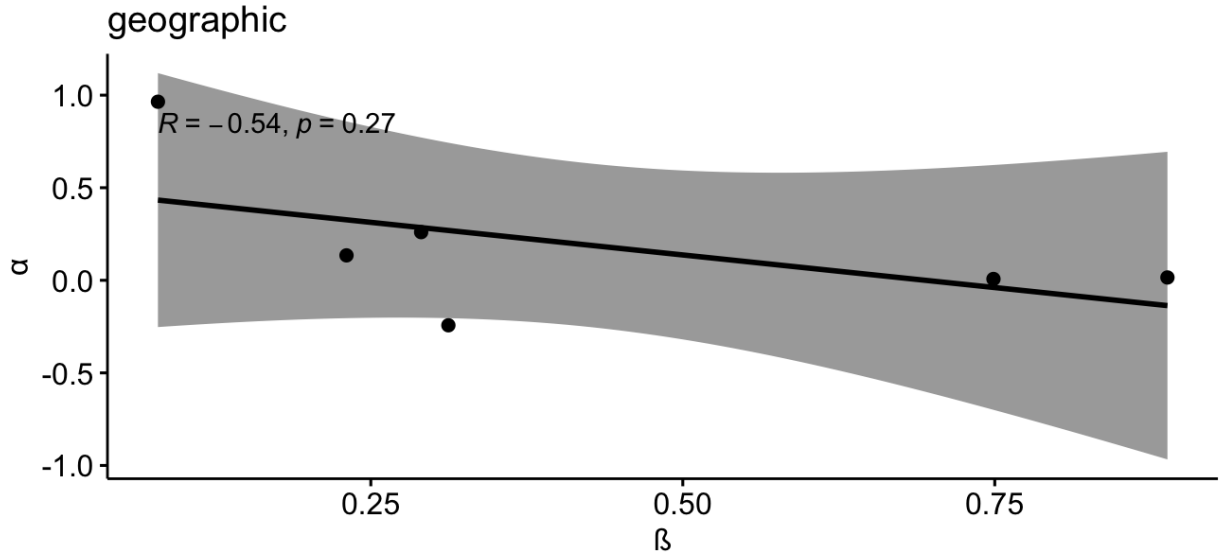

**Figure 8**: Using the ggscatter function, a correlation between the strong-weak ratio and the sentiment is highlighted. It has a correlation coefficient of -0.54 and a p-value of 0.27.

The correlation between ß and α for the Individual Query leaned towards the alternative hypothesis, $H_a$. And since $p > 0.01$, it is significant enough to provide evidence for the alternative hypothesis. In this case, the correlation between ß and α is significantly greater than for other queries.

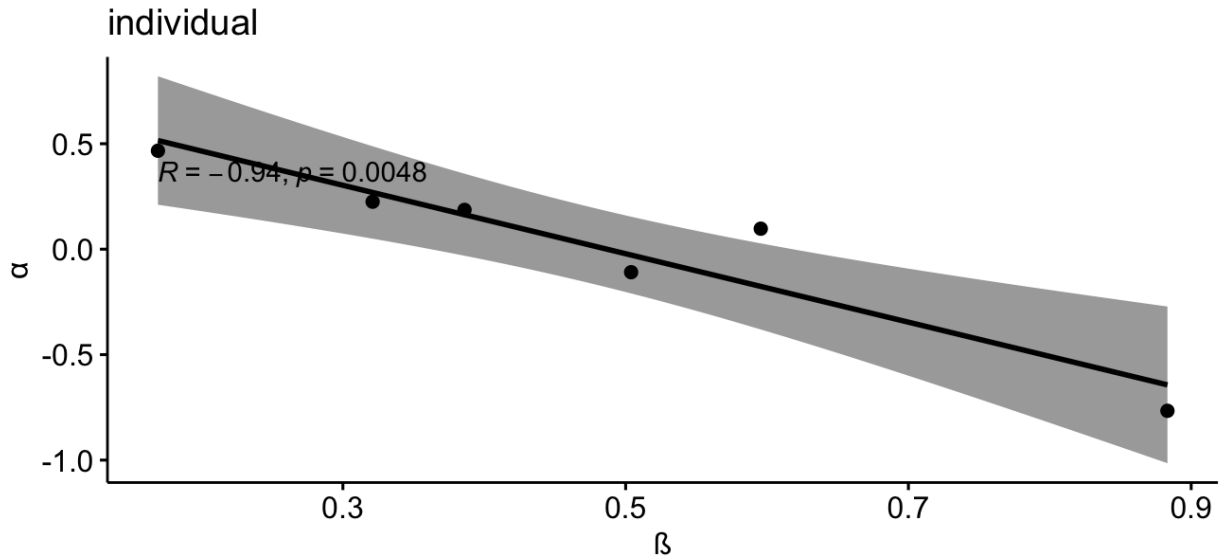

**Figure 9**: Using the ggscatter function, a correlation between the strong-weak ratio and the sentiment is highlighted. It has a correlation coefficient of -0.94 and a p-value of 0.0048.

Considering each of the six subjects underwent a hundred iterations and the pattern was still significant within the Individual Query, there may be a potentially inverse behavior between the closeness of an online community and the sentiment of its conversations.

### 4.2 Zou Interval and Fisher Score Comparisons for Cross-Query Data

In regards to the second set of hypotheses, the correlation coefficients between each query type can be analyzed in order to provide evidence for the alternative. Due to the random selection within the API's "mixed" strategy and the several iterations on our end, it is highly unlikely that one user was present in every sample. Additionally, considering the random selection present within the selection of the topics, it is unlikely that a user who is present in multiple of our query subjects behaves in the same manner across all. This can be supported by notions on dynamic behavior, Twitter API selection, and novel works which follow similar processes (Cont et al., 2000; Pfeffer et al., 2018). As such, we can treat each of the queries as an independent group. Using Zou (2007) and Fisher (1925)'s procedure to compare independent correlation coefficients across populations with z-scores, we can analyze if differences across query types exist.

With the "cocor" package in R (Diedenhofen et al., 2015), we can compare our correlation coefficients. In order to provide a more specific cross-query analysis, each query type is compared to another, instead of a simultaneous comparison between all four. Below is a table highlighting the results of each test of significance conducted for between-group variance.

| Groups | Z-Score | P-Value | Z-Score Hypothesis Result | Zou CI (95%) | Zou CI Hypothesis Result |
|---|---|---|---|---|---|
| Topic - Event | -12.6348 | 0 | Rejected | (-0.5102, -0.3654) | Rejected |
| Event - Geographic | 4.8058 | 0 | Rejected | (0.1219, 0.2903) | Rejected |
| Geographic - Individual | 21.5536 | 0 | Rejected | (0.3499, 0.4562) | Rejected |
| Topic - Geographic | -7.8018 | 0 | Rejected | (-0.2915, -0.1705) | Rejected |
| Event - Individual | 26.3594 | 0 | Rejected | (0.5424, 0.6751) | Rejected |
| Topic - Individual | 13.7518 | 0 | Rejected | (0.1413, 0.2037) | Rejected |

**Table 5**: The results of "cocor" analysis on each relationship combination within the query types is presented. The p-value is the likelihood of the z-score considering the second null hypothesis is true. The confidence interval (CI) provides a range of differences for the relationships highlighted.

Both confidence intervals tests and z-score tests highlight that the null hypothesis had to have been false for this data to be possible. As such, supporting $H_b$ and the notion of a difference between each query type.

## 5 Conclusions and Future Work

Our research has investigated two main concepts: (a) the correlation between structure and sentiment in online conversations, and (b) differences in this correlation for varying queries. Within these two subjects, we have found a general behavior for all queries which suggests that there is an inverse relationship between strong, close knit structure and sentiment. We also have found notable differences within this behavior for each query type. And although the behavior is not reversed within different queries, the degree of influence structure has in its relationship is highest for individual queries and lowest for event-based queries. Adoption of the majority-perspective and identity sharing (Dovidio et al., 2010) may be responsible for the peak seen with individual queries, A possible explanation for the minimum with event-based queries can come from Zacks et al. (2007) where perpetually dynamic perspectives and event segmentation cause diverse opinions on one event. With further research, we can possibly provide a factorable relationship between structure and sentiment. With this, we can analyze at-risk communities and warn them of polarization, develop tools to protect electoral processes, and provide effective technological moderation when needed. However, there are currently no quantitative methods to prove a causal relationship between structure and sentiment for any of these queries. And this work aims to spark an initiative to further explore this realm.

Whether it be in-person or online, experiments and studies which involve studying sentiment must be aware of the influence of perspectives and how, even in large groups, opinion is often manipulated (Orléan, 1999). Due to the gravity of perspective, background issues such as differences within query types must be brought forth and addressed—in an effort to optimize our time for the bigger problems. This paper aims to make researchers aware of these differences and hopes the methods highlighted can help further studies to develop tools and ideas for social exploration.